\documentclass[10pt]{article}
\usepackage{authblk}
\usepackage{amsmath}
\usepackage{amssymb}
\usepackage[left=10mm, top=10mm, right=10mm, bottom=12mm]{geometry}

\usepackage{graphicx}
\usepackage{subfig}

\usepackage[font=small]{caption}

\begin{document}
\title{Three-dimensional static black hole with $\Lambda$ and nonlinear electromagnetic fields and its thermodynamics}
\author{M. B. Tataryn\footnote{E-mail: misha.physics@gmail.com}, M. M. Stetsko\footnote{E-mail: mstetsko@gmail.com}}
\affil{\small Department for Theoretical Physics,\\Ivan Franko National University of Lviv,\\12 Drahomanov Str., Lviv, 79005, Ukraine}
\date{}
\maketitle
\abstract{Static black hole with the Power Maxwell invariant (PMI), Born-Infeld (BI), logarithmic (LN), exponential (EN) electromagnetic fields in three-dimensional spacetime with cosmological constant was studied. It was shown that the LN and EN fields represent the Born-Infeld type of nonlinear electrodynamics. It the framework of General Relativity the exact solutions of the field equations were obtained, corresponding thermodynamic functions were calculated and the $P-V$ criticality of the black holes in the extended phase space thermodynamics was investigated.

}
\section{Introduction}
General Relativity gives excellent results in description of macroscopic gravitating objects in the Universe. One of the types of objects that need such a description are black holes, the characteristic and distinguishing feature of which is the existence of the so-called event horizon. The interest in black hole physics is caused by their extremal (specific) physical characteristics and behavior. There are a lot of works on the physics of black holes, static and rotating, electrically neutral and charged, with a cosmological constant and without it. Thermodynamics of black holes is also area of active investigation. There are numerous works on black holes in spacetime of various dimensions, on black holes with nonlinear electromagnetic fields. The works where gravity is coupled to linear Maxwell field can be found for example in \cite{bel12}. One of the special types of nonlinear fields are the PMI in (2+1) \cite{hen14}, (3+1) \cite{pan19}, (n+1) \cite{has08,gon09,hen13} and the Born-Infeld fields in (3+1) \cite{fer03,stef07,myun08,chem08,ban12} and (n+1) \cite{dey04,cai04,mis08,gunas12,zou14} cases. The specific case of the PMI field is $s=d/4$, where $d=n+1$ is the dimension of spacetime, wherein is achieved so-called conformal invariance for one of the types of the conformal transformations (conformal field theory, CFT).

Some attention is also devoted to exponential field in (3+1) \cite{shey14} and exponential and logarithmic fields in (n+1) \cite{hen15} spacetime. Besides it the Heisenberg-Euler-type electrodynamics was considered \cite{krug17}. Especially the black holes thermodynamics on extended phase space was investigated and $P-V$ criticality was studied. Black hole thermodynamics with $P-V$ term was considered in \cite{kubiz12,kubiz17}. It was obtained exact solutions of field equations, considered thermodynamic quantities, verified the first law of black hole thermodynamics in the extended phase space thermodynamics. Besides of all that, there are works on modified theory of gravity, for example, $F(R)$ gravity \cite{hen14}, scalar-tensor theories \cite{stef07}, theories where gravity is coupled to dilaton \cite{shey14,hen18}. Also it was studied the magnetically charged black hole in framework of nonlinear electrodynamics \cite{krug18}. 

The interest in studying of three-dimensional black holes (BTZ) \cite{ban92} and in general three-dimensional gravity theories is due to many reasons. It was shown that three-dimensional black holes possess all the specific features of four and higher dimensional black holes, in particular it was demonstrated in the work \cite{ban92} and other authors (for example, see \cite{hen14}), but due to lower dimensional three-dimensional black holes are relatively simple in comparison with higher dimensional ones. It is known that in three-dimensional case black holes solutions appear when the negative cosmological constant is taken into account, thus their examination might be interesting due to AdS/CFT correspondence. In this case AdS/CFT correspondence allows to find relations between three-dimensional black holes and two-dimensional conformal field theory and as a result (2+1)-dimensional black holes give new opportunities for studying conformal field theory problems and vice versa \cite{hen18,wit}. Some other aspects of three-dimensional black holes were also studied, namely scale dependent black holes with nonlinear field was obtained and scalar quasinormal modes were calculated \cite{rin17,rin18_2,rin18}. Propagation of a probe minimally coupled scalar field in black hole with PMI source for three dimensions was also considered \cite{pan18}.

Taking into account the well-studied black hole solutions in presence of nonlinear electromagnetic source it seems useful to us to consider a few different nonlinear field types with their following comparison and study how these different fields affect the thermodynamics of a black hole. For this purpose we choose such fields like the more studied PMI and Born-Infeld fields and less investigated logarithmic and exponential fields applied to the static (2+1) black hole system. The static case is chosen mainly for simplicity and the consideration of three-dimensinal gravity is caused by mentioned above characteristic features of three-dimensional black holes. We think that such formulation of the problem which includes the comparison of different field types will help to systematize the already available results. Thus, this work have self-sufficient and complete character.

The outline of current paper is as follows. We examine a charged static black hole with four different kinds of a nonlinear electromagnetic fields and cosmological constant in a three-dimensional (2+1) spacetime, namely we consider the PMI (Power Maxwell invariant), Born-Infeld, logarithmic and exponential electromagnetic fields. We start with an expression for the functional of action of black hole system and an expression for invariant interval. In the framework of General Relativity we obtain equations for gravitation and electromagnetic fields and their exact solutions. On their basis we calculate the temperature and electric potential on the event horizon of black hole. Then we examine so-called mass function of black hole and the first law of black hole thermodynamics. After this we consider the cosmological constant as one of the thermodynamic quantities, namely pressure. According to this we obtain one additional term in the first law, namely pressure-volume term and calculate thermodynamic volume. We obtain equations of state and calculate the heat capacity. We investigate the behavior of the results and finish our article with some conclusions.
\section{Field equations and their solutions}
We are going to obtain black hole solutions with nonlinear electromagnetic field and cosmological constant in a three-dimensional (2+1) spacetime.  The action in this case can be written as follows ($G\equiv1$) \cite{hen14}
\begin{equation}
I(g_{\mu\nu},A_\mu)=\frac{1}{16\pi}\int\limits_\Sigma d^3x\sqrt{-g}[R-2\Lambda+L(F)],
\end{equation}
where $g_{\mu\nu}$ is the metric tensor, $A_\mu$ is the electromagnetic potential, $g\equiv\det||g_{\mu\nu}||$, $R$ is the scalar curvature, $\Lambda$ is the negative (AdS black hole) cosmological constant, $L(F)$ is the lagrangian of electromagnetic field, $F=F_{\mu\nu}F^{\mu\nu}$ is the Maxwell invariant, $F_{\mu\nu}$ is the electromagnetic tensor, $F_{\mu\nu}=\partial_\mu A_\nu-\partial_\nu A_\mu$, where $\partial_\mu\equiv\partial/\partial x^{\mu}$. We consider four kinds of nonlinear electromagnetic fields \cite{hen14,shey14}: PMI (Power Maxwell invariant) $L_{PMI}=(-F)^s$, Born-Infeld field $L_{BI}=4\beta^2\left(1-\sqrt{1+\frac{F}{2\beta^2}}\right)$, logarithmic field $L_{LN}=-4\beta^2\ln\left(1+\frac{F}{4\beta^2}\right)$ and exponential field $L_{EN}=4\beta^2\left(e^{-\frac{F}{4\beta^2}}-1\right)$, where $s$ and $\beta$ are parameters of these fields. When $s=1$ (PMI) and $\beta\to+\infty$ (BI, LN, EN) we obtain $L_{PMI}=L_{BI}=L_{LN}=L_{EN}=-F$, that is a linear Maxwell electromagnetic field. We obtain equations for gravitation field
\begin{equation}
R_{\mu\nu}-\frac{1}{2}g_{\mu\nu}R+g_{\mu\nu}\Lambda=T_{\mu\nu},
\end{equation}
where $R_{\mu\nu}$ is the Ricci curvature tensor, $T_{\mu\nu}$ is the stress-energy tensor
\begin{equation}
T_{\mu\nu}=\frac{1}{2}g_{\mu\nu}L(F)-2\frac{\partial L(F)}{\partial F}F_{\mu\beta}F_\nu^{\ \beta}.
\end{equation}
The equations for electromagnetic field
\begin{equation}
\partial_\mu\left(\sqrt{-g}\frac{\partial L(F)}{\partial F}F^{\mu\nu}\right)=0.
\end{equation}

We consider a three-dimensional (2+1) spacetime, $x^0=t$, $x^1=r$, $x^2=\varphi$, where $t$ is the time coordinate ($c\equiv1$), $r$ is the radial coordinate, $\varphi$ is the polar angle coordinate. In our work we study a static solution and the line element (metric) takes the form \cite{hen14}
\begin{equation}
ds^2=-g(r)dt^2+\frac{dr^2}{g(r)}+r^2d\varphi^2,
\end{equation}
where $g(r)$ is the metric function. For our metric \cite{hen14} $A_\mu=0$, $\mu\neq0$, $A_0=A_0(r)$, so only $F_{10}(r)=-F_{01}(r)\ne0$, $F_{10}=dA_0(r)/dr$. Here it is also necessary to note that as we mentioned in the introduction, the PMI case with $s=d/4$ is invariant with respect to so-called conformal transformations and for our three-dimensional spacetime we have that for $s=3/4$ the action becomes invariant with scale transformations as $r\to\lambda r$, $t\to\lambda t$, where $\lambda$ is the scale parameter.

We obtain exact solutions for the metric function $g(r)$ and the radial component of electric field $F_{10}(r)$ and they are shown in the Table~\ref{tab_g(r)} and Table~\ref{tab_F(r)}, respectively. 
\begin{table}[h!]
\begin{center}
\begin{tabular}{|ll|}
\hline
$g(r)=-\Lambda r^2-2q^2\ln\left(\frac{r}{l}\right)-m$ & PMI, $s=1$ \\
$g(r)=-\Lambda r^2-\frac{2^{s-1}(2s-1)^2}{s-1}q^{\frac{2s}{2s-1}}r^{\frac{2(s-1)}{2s-1}}-m$ & PMI, $s\ne1$ \\
$g(r)=(2\beta^2-\Lambda)r^2-2q^2\operatorname{arsinh}\left(\frac{\beta r}{q}\right)-2\beta q r\sqrt{1+\frac{\beta^2r^2}{q^2}}-m$ & BI \\
$g(r)=(3\beta^2-\Lambda)r^2+(2\beta^2r^2-2q^2)\operatorname{arsinh}\left(\frac{\beta r}{\sqrt{2}q}\right)-2\beta^2r^2\ln\left(\frac{\sqrt{2}\beta r}{q}\right)-3\sqrt{2}\beta qr\sqrt{1+\frac{\beta^2r^2}{2q^2}}-m$ & LN \\
$g(r)=-(\Lambda+2\beta^2)r^2+2q^2\left(\frac{1}{W}-2\right)e^{-\frac{W}{2}}-q^2\operatorname{Ei}\left(\frac{W}{2}\right)-m$ & EN \\
\hline
\end{tabular}
\caption{Metric function $g(r)$.}
\label{tab_g(r)}
\end{center}
\end{table}
\begin{table}[h!]
\begin{center}
\begin{tabular}{|ll|}
\hline
$F_{10}(r)=\left(\frac{q}{r}\right)^{\frac{1}{2s-1}}$ & PMI \\
$F_{10}(r)=\frac{q}{\sqrt{\frac{q^2}{\beta^2}+r^2}}$ & BI \\
$F_{10}(r)=-\frac{\beta^2r}{q}+\sqrt{\frac{\beta^4r^2}{q^2}+2\beta^2}$ & LN \\
$F_{10}(r)=\beta\sqrt{W}$ & EN \\
\hline
\end{tabular}
\caption{Radial component of electric field $F_{10}(r)$.}
\label{tab_F(r)}
\end{center}
\end{table}

In the tables we have $l$ is constant, $[l]=[r]$, $q$ and $m$ are constants of integration, they are related to the charge and mass of the black hole, respectively. $W=W\left(\frac{q^2}{\beta^2r^2}\right)$, $W(x)$ is the Lambert function, $W(x)e^{W(x)}=x$, $\operatorname{Ei}(x)$ is the exponential integral, $\operatorname{Ei}(x)=\int\limits_1^{+\infty}\frac{e^{-xt}}{t}dt$. As it can be seen from the Table~\ref{tab_g(r)}, for all field types the term with the cosmological constant becomes dominating for large $r$ and we have the quadratic dependence for $g(r)$. For small $r$ the effect of $\Lambda$ term on $g(r)$ is negligibly small and here the other terms are more significant, for example, they give a minimum of the metric function. Since the dependence of the metric function $g(r)$ of $m$ is linear, it means that any change of the mass parameter leads to vertical shift of corresponding graphics for the function $g(r)$ and it affects the existence of the event horizon. The increasing of the absolute value of the cosmological constant $\Lambda$ gives rise to faster increase of the function $g(r)$ for large $r$ and this reduces the value of the event horizon. The increase of the charge parameter $q$ leads to corresponding increase of the radius of the event horizon of the black hole, what is opposite to the situation mentioned above where the increase of the absolute value of $\Lambda$ leads to decrease of the horizon's radius. The difference is manifested only depenging the metric function $g(r)$ on parameter $q$ for PMI field for $s<1$, as in this case the term with $q$ takes the positive sign. For example when $s=3/4$ we obtain that the increasing of the $q$ gives rise to faster increase of the $g(r)$ for large $r$.

\begin{figure}[h!]
\centering
\includegraphics[width=0.4\textwidth]{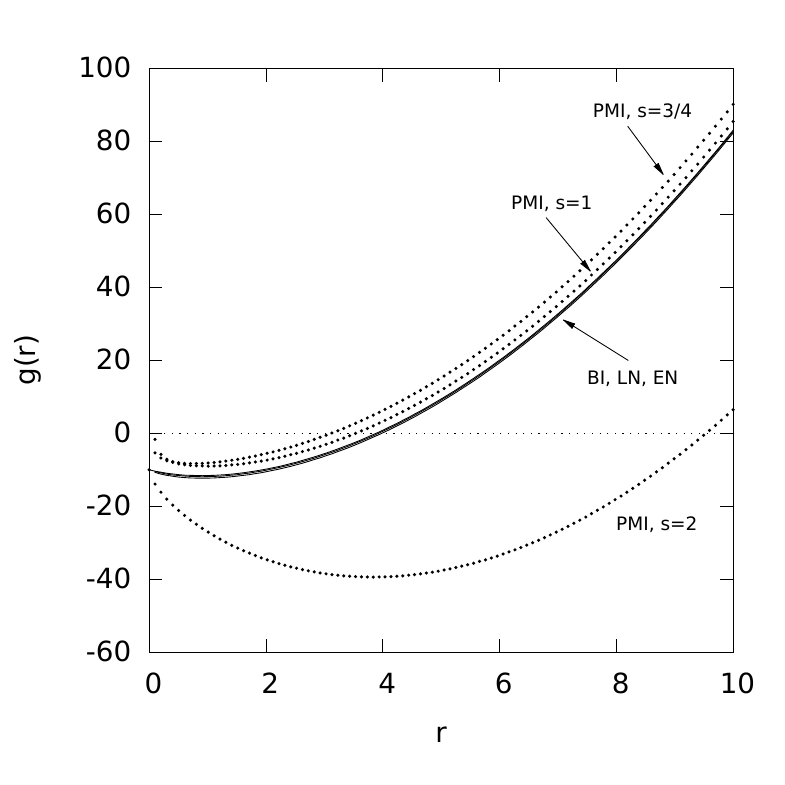}
\caption{The metric function $g(r)$ for $l=1$, $\beta=1$, $\Lambda=-1$, $q=1$, $m=10$.}
\label{fig_g(r)}
\end{figure}

\begin{figure}[h!]
\centering
\subfloat[different $\Lambda$]{\includegraphics[width=0.3\columnwidth]{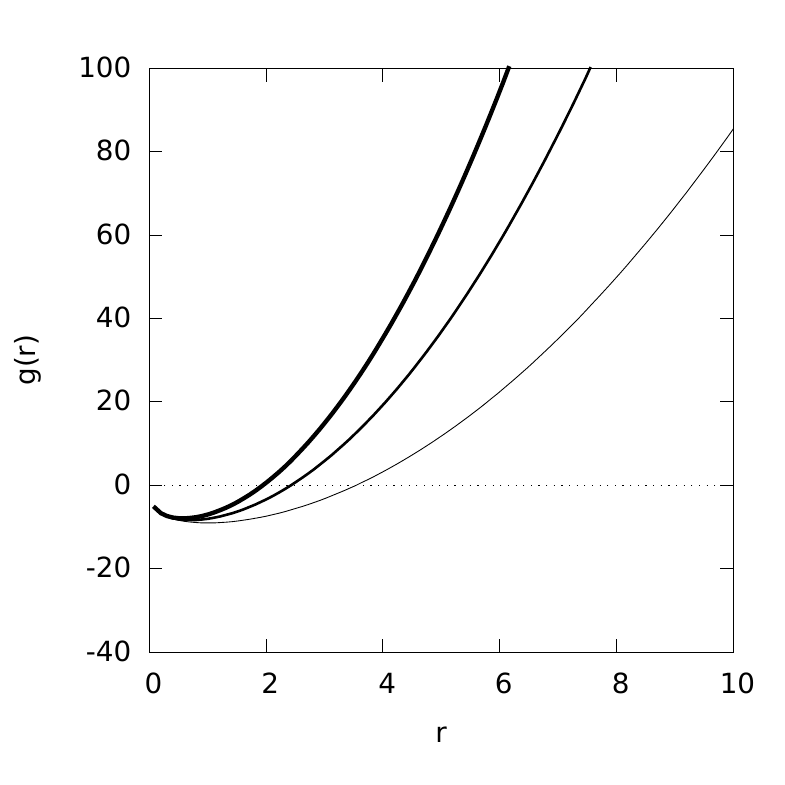}}
\qquad
\subfloat[different $q$]{\includegraphics[width=0.3\columnwidth]{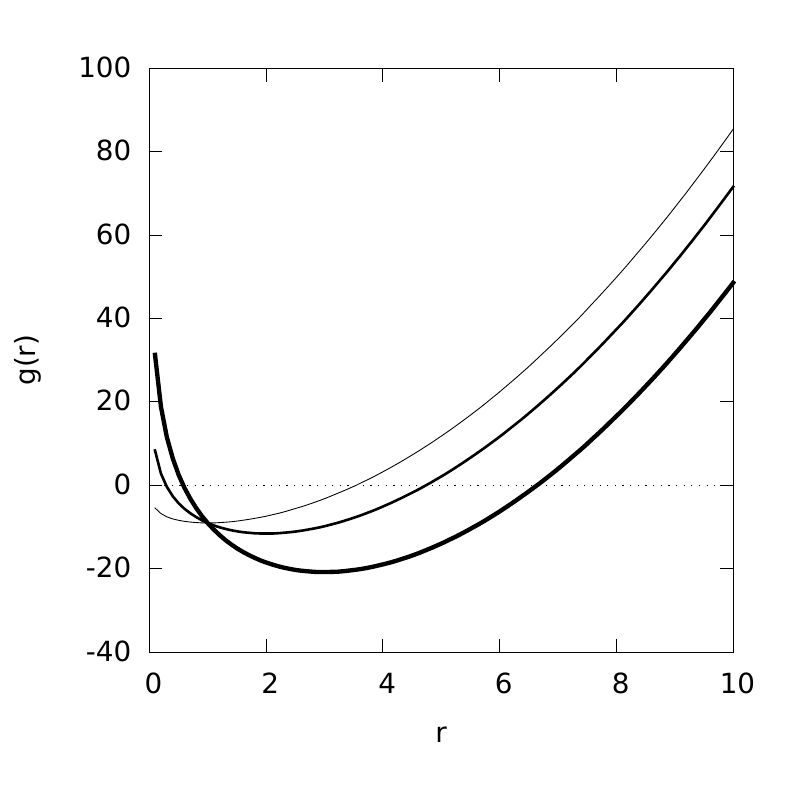}}
\caption{$g(r)$ for linear field (s=1) with $l=1$, $m=10$, (a) $\Lambda=-1$ (thin), $\Lambda=-2$ (medium), $\Lambda=-3$ (bold), $q=1$, (b) $q=1$ (thin), $q=2$ (medium), $q=3$ (bold), $\Lambda=-1$.}
\label{fig_g(r)_PMI_s_1_Lambda_q}
\end{figure}

The graphs of $g(r)$ for all kinds of fields with the same fixed values of other parameters are presented on Fig.~\ref{fig_g(r)}. Their general behavior is similar and for BI, LN and EN fields we have almost identical lines. The behavior of curves for the PMI depends on $s$, and for $s=1$ and $s=3/4$ the curves are very close to BI, LN, EN cases, whereas for $s=2$, the event horizon is considerably greater than in the mentioned above cases. On Fig.~\ref{fig_g(r)_PMI_s_1_Lambda_q} the metric function $g(r)$ for linear field is shown for different values of $\Lambda$ (the left graph) and $q$ (the right one).

In Table~\ref{tab_F(r)} we have solutions for radial component of electric field $F_{10}(r)$ for all field types. These dependences are monotonous decreasing functions on $r$. The difference is manifested only for the PMI when $s<1/2$, where in this case we obtain unlimited increase of $F_{10}(r)$ for large $r$ and also $F_{10}(r)\to0$  when $r\to0$. Below we do not consider the PMI case of $s<1/2$. The increase of charge parameter $q$ leads to faster rise of $F_{10}(r)$ when $r\to0$.

\begin{figure}[h!]
\centering
\includegraphics[width=0.4\textwidth]{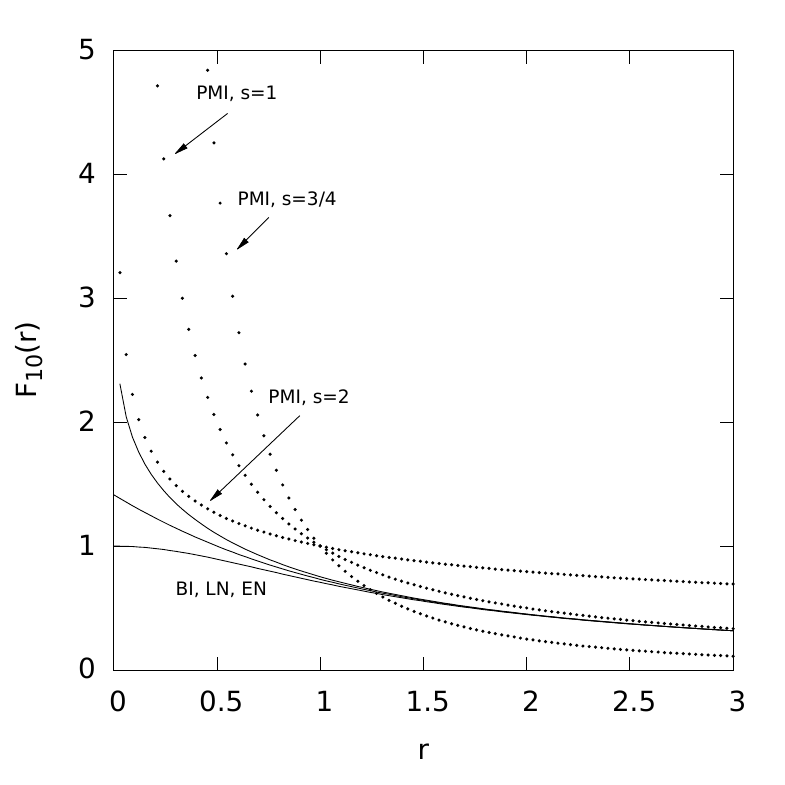}
\caption{The radial component of electric field $F_{10}(r)$ for $\beta=1$, $q=1$.}
\label{fig_F(r)}
\end{figure}

\begin{figure}[h!]
\centering
\subfloat[different $q$ for linear field]{\includegraphics[width=0.3\columnwidth]{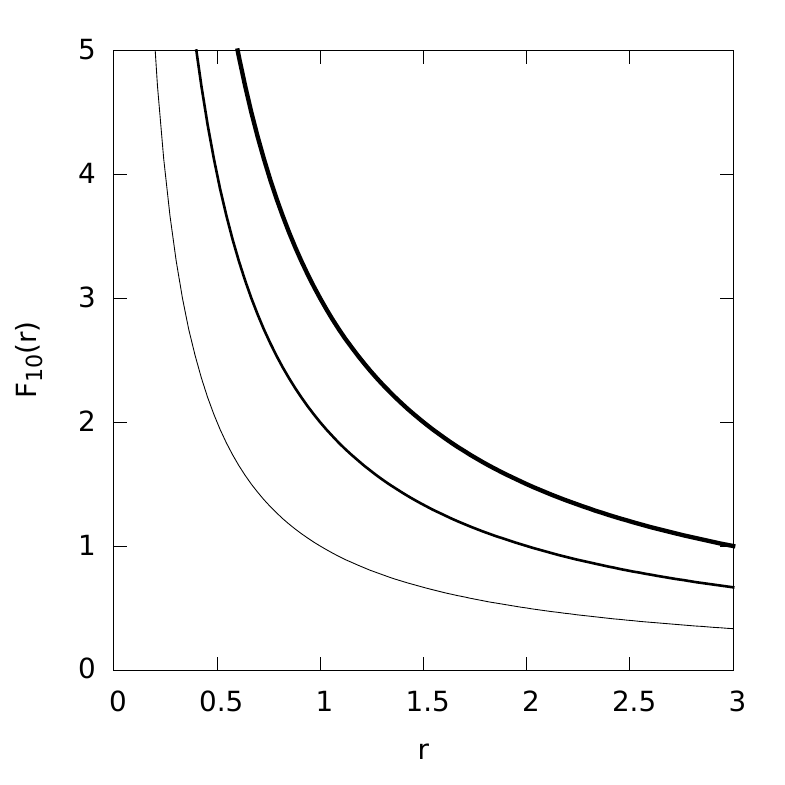}}
\qquad
\subfloat[different $\beta$ for BI field]{\includegraphics[width=0.3\columnwidth]{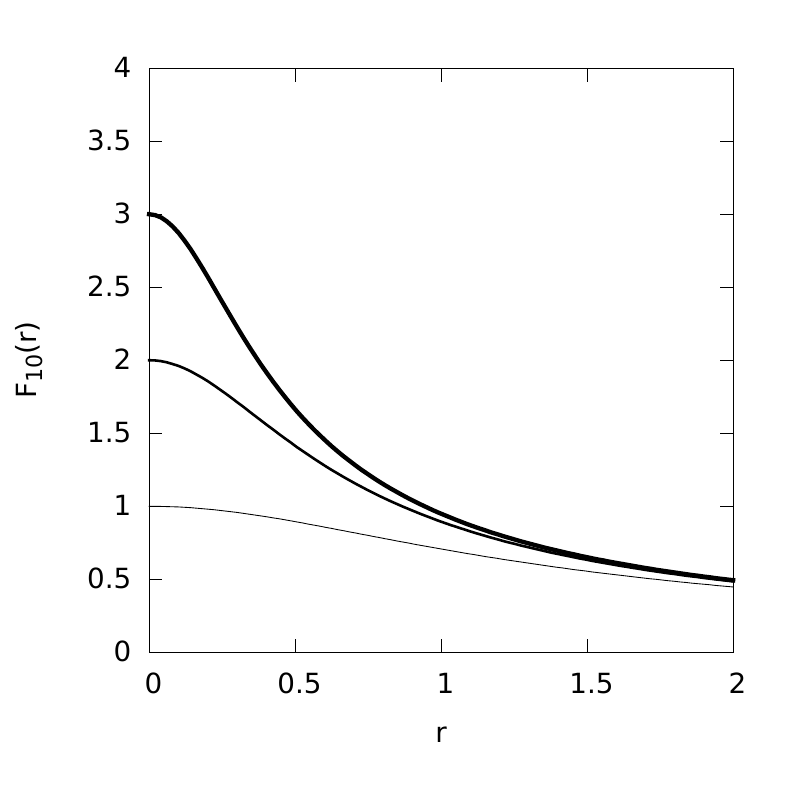}}
\caption{$F_{10}(r)$ for (a) linear field with $q=1$ (thin), $q=2$ (medium), $q=3$ (bold), (b) BI field with $\beta=1$ (thin), $\beta=2$ (medium), $\beta=3$ (bold), $q=1$.}
\label{fig_F(r)_PMI_s_1_q_BI_beta}
\end{figure}

On Fig.~\ref{fig_F(r)} the function $F_{10}(r)$ are presented for all kinds of nonlinear field with the fixed values of charge parameter $q$. For large distances we obtain for BI, LN and EN fields almost identical curves as for linear field (PMI, $s=1$). The BI and LN fields at small ($r\to0$) values of $r$ gives limited electric field $F_{10}(r)$, whereas for EN and PMI fields the function $F_{10}(r)$ goes to infinity. On Fig.~\ref{fig_F(r)_PMI_s_1_q_BI_beta} the function $F_{10}(r)$ is shown for linear field with different values of $q$ $(a)$ and for BI field for different values of $\beta$ $(b)$.
\section{Thermodynamics}
\subsection{Themperature and electric potential}
Let $r=r_+$ is the event horizon of black hole, $g(r_+)=0$. Will consider the temperature $T$ and electric potential $U$ on the event horizon. The Hawking themperature can be written as \cite{hen14}
\begin{equation}
T=\left.\frac{1}{2\pi}\sqrt{-\frac{1}{2}(\nabla_\mu\chi_\nu)(\nabla^\mu\chi^\nu)}\right|_{r=r_+}
\end{equation}
and electric potential in the form
\begin{equation}
U=\left.-sA_\mu\chi^\mu\right|_{r=r_+}+C
\end{equation}
for PMI, and in the form
\begin{equation}
U=\left.-A_\mu\chi^\mu\right|_{r=r_+}+C
\end{equation}
for BI, LN and EN, where $\nabla_\mu$ is covariant derivative, $\chi^\mu$ is a timelike Killing vector field, we have $\chi^\mu=(1,0,0)$. We obtain $T=\left.\frac{1}{4\pi}\frac{dg(r)}{dr}\right|_{r=r_+}$, $U=\left.-sA_0\right|_{r=r_+}+C$ for PMI and $U=\left.-A_0\right|_{r=r_+}+C$ for BI, LN and EN, where $A_0=\int F_{10}dr+C$, $C$ is a constant of integration. We obtain formulas for the temperature (Table~\ref{tab_T(r)}) and electric potential (Table~\ref{tab_U(r)}), where $W=W\left(\frac{q^2}{\beta^2r_+^2}\right)$. 
\begin{table}[h!]
\begin{center}
\begin{tabular}{|ll|}
\hline
$T=-\frac{1}{2\pi}\left(\Lambda r_++2^{s-1}(2s-1)q^{\frac{2s}{2s-1}}r_+^{-\frac{1}{2s-1}}\right)$ & PMI \\
$T=-\frac{1}{2\pi}\left((\Lambda-2\beta^2)r_++2\beta q\sqrt{1+\frac{\beta^2r_+^2}{q^2}}\right)$ & BI \\
$T=-\frac{1}{2\pi}\left((\Lambda-2\beta^2)r_+-2\beta^2r_+\operatorname{arsinh}\left(\frac{\beta r_+}{\sqrt{2}q}\right)+2\beta^2r_+\ln\left(\frac{\sqrt{2}\beta r_+}{q}\right)+2\sqrt{2}\beta q\sqrt{1+\frac{\beta^2r_+^2}{2q^2}}\right)$ & LN \\
$T=-\frac{1}{2\pi}\left((\Lambda+2\beta^2)r_++2\beta q(\sqrt{W}-W^{-\frac{1}{2}})\right)$ & EN \\
\hline
\end{tabular}
\caption{The Hawking temperature $T(r_+)$.}
\label{tab_T(r)}
\end{center}
\end{table}
\begin{table}[h!]
\begin{center}
\begin{tabular}{|ll|}
\hline
$U=-q\ln\left(\frac{r_+}{l}\right)$ & PMI, $s=1$ \\
$U=-\frac{s(2s-1)}{2(s-1)}q^{\frac{1}{2s-1}}r_+^{\frac{2(s-1)}{2s-1}}$ & PMI, $s\neq1$ \\
$U=-q\operatorname{arsinh}\left(\frac{\beta r_+}{q}\right)$ & BI \\
$U=\frac{\beta^2r_+^2}{2q}-q\operatorname{arsinh}\left(\frac{\beta r_+}{\sqrt{2}q}\right)-\frac{\beta r_+}{\sqrt{2}}\sqrt{1+\frac{\beta^2r_+^2}{2q^2}}$ & LN \\
$U=-qe^{-\frac{W}{2}}-\frac{q}{2}\operatorname{Ei}\left(\frac{W}{2}\right)$ & EN \\
\hline
\end{tabular}
\caption{Electric potential $U(r_+)$.}
\label{tab_U(r)}
\end{center}
\end{table}

The temperature $T(r_+)$ are monotonous increasing functions for all field types and demonstrates similar qualitative general behavior. For small values of $r_+$ it takes negative values, but it can be shown that it takes place in nonphysical region. The dependences $T(r_+$) on parameters $\Lambda$ and $q$ is the same as for $g(r)$.

\begin{figure}[h!]
\centering
\includegraphics[width=0.4\textwidth]{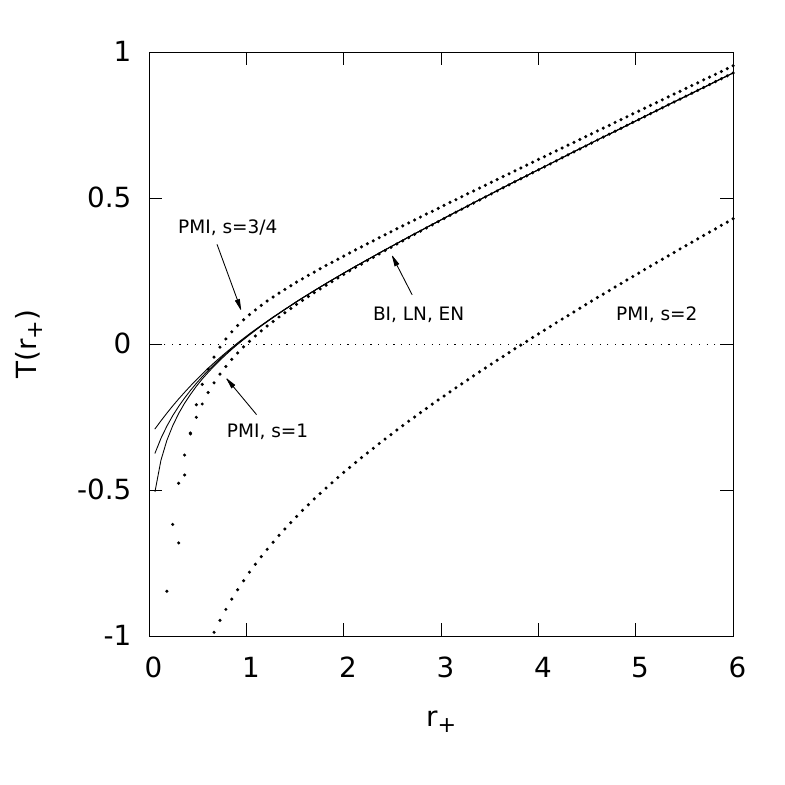}
\caption{The temperature $T(r_+)$ for $\beta=1$, $\Lambda=-1$, $q=1$.}
\label{fig_T(r)}
\end{figure}

\begin{figure}[h!]
\centering
\subfloat[different $\Lambda$]{\includegraphics[width=0.3\columnwidth]{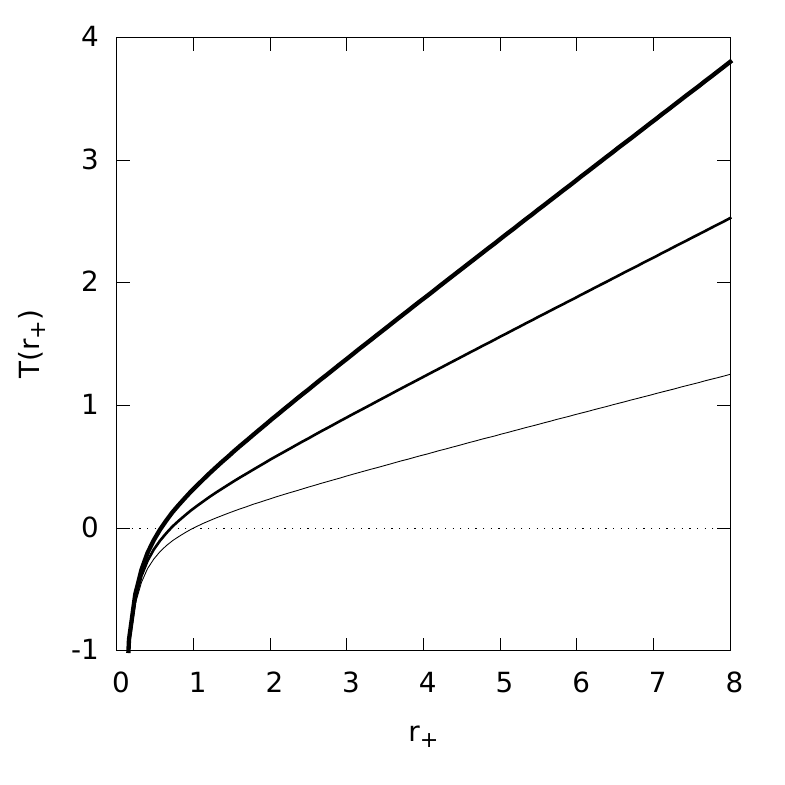}}
\qquad
\subfloat[different $q$]{\includegraphics[width=0.3\columnwidth]{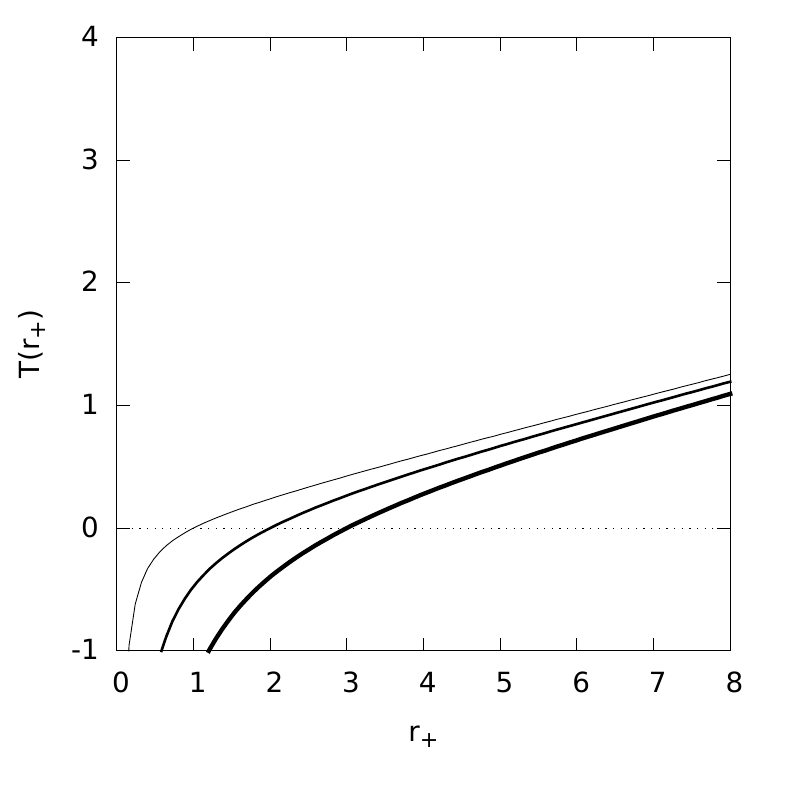}}
\caption{$T(r_+)$ for linear field with (a) $\Lambda=-1$ (thin), $\Lambda=-2$ (medium), $\Lambda=-3$ (bold), $q=1$, (b) $q=1$ (thin), $q=2$ (medium), $q=3$ (bold), $\Lambda=-1$.}
\label{fig_T(r)_PMI_s_1_Lambda_q}
\end{figure}

The graphs for $T(r_+)$ are presented on Fig.~\ref{fig_T(r)}. For BI, LN and EN cases we get again almost identical lines and difference between other curves is the same as for $g(r)$ on Fig.~\ref{fig_g(r)}. On Fig.~\ref{fig_T(r)_PMI_s_1_Lambda_q} the temperature $T(r_+)$ is shown depending on $\Lambda$ and $q$ for linear field.
\subsection{The first law of black hole thermodynamics}
Now we consider so-called mass function of black hole \cite{hen14}
\begin{equation}
M=\frac{m(r_+, q)}{8},
\end{equation}
where $m$, $q$ -- the constants of integration. We know that we have linear homogeneous functions $r_+=r_+(S)$ and $q=q(Q)$, where $S$ is the entropy of the black hole, $Q$ is the electric charge of the black hole. $S=A/4$, where $A$ is the area of the event horizon, we have $A=\int\limits_0^{2\pi}r_+d\varphi=2\pi r_+$ and we obtain $r_+=2S/\pi$. The relation between $q$ and $Q$ can be given by the Gauss's law, we obtain $q=2^{2-s}Q$ (PMI), $q=2Q$ (BI, LN, EN). So, we have the mass function $M(S,Q)$ and
\begin{equation}
dM=TdS+UdQ.
\end{equation}
This is the first law of black hole thermodynamics. Temperature and electric potential are equal to $T=\left(\frac{\partial M}{\partial S}\right)_Q$ and $U=\left(\frac{\partial M}{\partial Q}\right)_S$, respectively. These formulas for $T$ and $U$ give the same results as we got previously in variables $r_+$, $q$ (Table~\ref{tab_T(r)}, Table~\ref{tab_U(r)}).
\subsection{Pressure-volume term and equations of state}
Let us consider the cosmological constant $\Lambda$ as one of the thermodynamic quantities, it was shown that it is related to the pressure \cite{kubiz17}
\begin{equation}
P=-\frac{\Lambda}{8\pi}.
\end{equation}
Now the mass function $M(S,Q,P)$ is identified with enthalpy and we can write
\begin{equation}
dM=TdS+UdQ+VdP.
\end{equation}
We obtain the thermodynamic volume $V=\left(\frac{\partial M}{\partial P}\right)_{S,Q}$ for all kinds of fields in  the form
\begin{equation}
V=\pi r_+^2,
\end{equation}
or $V=4S^2/\pi$ in thermodynamic variables. We can also write the analogous expressions for $T$ and $U$ in thermodynamic quantities $S$, $Q$, $P$, they are in linear homogeneous relations with $r_+$, $q$, $\Lambda$, respectively and we do not write them again. We obtain the equations of state $f(P,V,T,Q)$ (Table~\ref{tab_f(PVTQ)}), where $W=W\left(\frac{4\pi Q^2}{\beta^2V}\right)$.
\begin{table}[h!]
\begin{center}
\begin{tabular}{|ll|}
\hline
$P=\frac{\sqrt{\pi}T}{4\sqrt{V}}+2^{-\frac{5s-4}{2s-1}}\pi^{-\frac{s-1}{2s-1}}(2s-1)Q^{\frac{2s}{2s-1}}V^{-\frac{s}{2s-1}}$ & PMI \\
$P=\frac{\sqrt{\pi}T}{4\sqrt{V}}+\frac{\beta Q}{2\sqrt{\pi V}}\sqrt{1+\frac{\beta^2 V}{4\pi Q^2}}-\frac{\beta^2}{4\pi}$ & BI \\
$P=\frac{\sqrt{\pi}T}{4\sqrt{V}}-\frac{\beta^2}{4\pi}\operatorname{arsinh}\left(\frac{\beta\sqrt{V}}{2\sqrt{2\pi}Q}\right)+\frac{\beta^2}{4\pi}\ln\left(\frac{\beta\sqrt{V}}{\sqrt{2\pi}Q}\right)+\frac{\beta Q}{\sqrt{2\pi V}}\sqrt{1+\frac{\beta^2V}{8\pi Q^2}}-\frac{\beta^2}{4\pi}$ & LN \\
$P=\frac{\sqrt{\pi}T}{4\sqrt{V}}+\frac{\beta Q\sqrt{W}}{2\sqrt{\pi V}}-\frac{\beta Q}{2\sqrt{\pi VW}}+\frac{\beta^2}{4\pi}$ & EN \\
\hline
\end{tabular}
\caption{Equations of state $f(P,V,T,Q)$.}
\label{tab_f(PVTQ)}
\end{center}
\end{table}

The functions $P(V)$ (for fixed parameters $T$ and $Q$) decrease monotonously with rise of $V$ and $P(V)\to+\infty$ when $V\to0$ for all field types. The change of parameters $T$ and $Q$ in the same way affects the dependences $P(V)$, namely the corresponding isotherm by large $T$ (or $Q$) is higher and to the right than for smaller values of $T$ (or $Q$).

\begin{figure}[h!]
\centering
\includegraphics[width=0.4\textwidth]{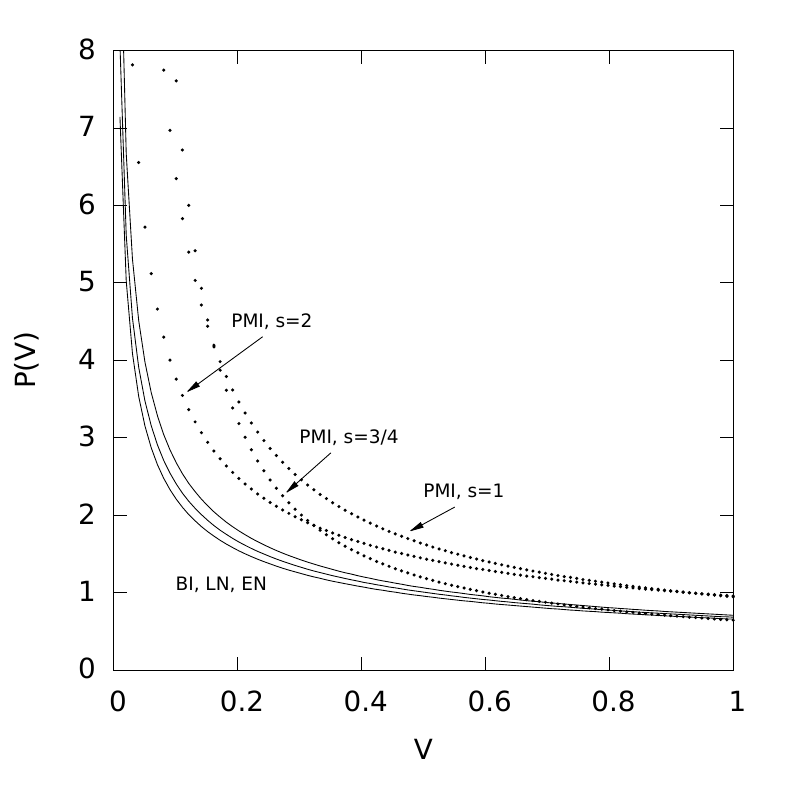}
\caption{$P(V)$ for $\beta=1$, $T=1$, $Q=1$.}
\label{fig_P(V)}
\end{figure}

\begin{figure}[h!]
\centering
\subfloat[different $T$]{\includegraphics[width=0.3\columnwidth]{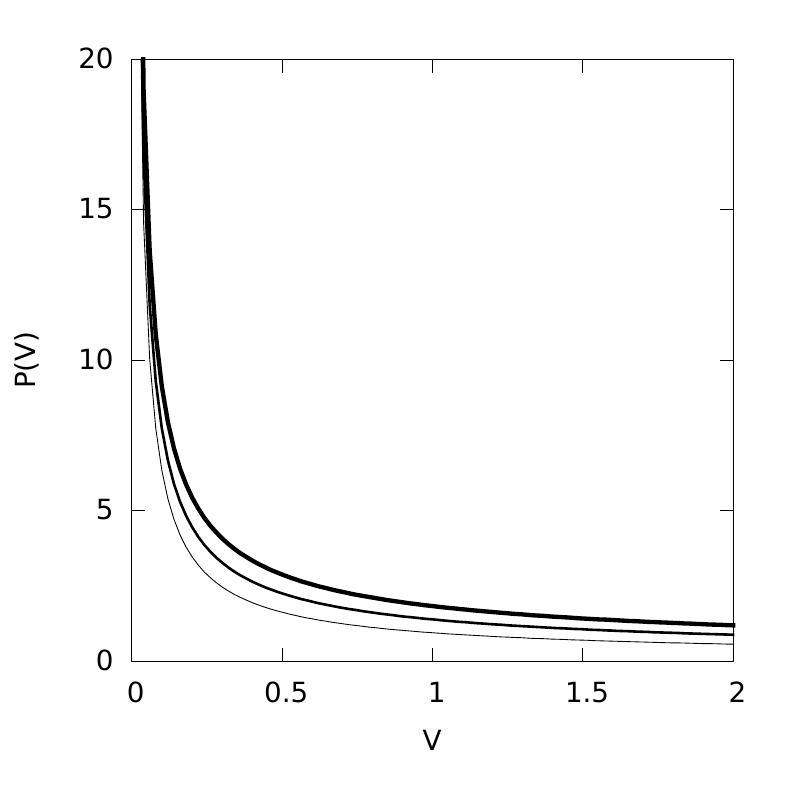}}
\qquad
\subfloat[different $Q$]{\includegraphics[width=0.3\columnwidth]{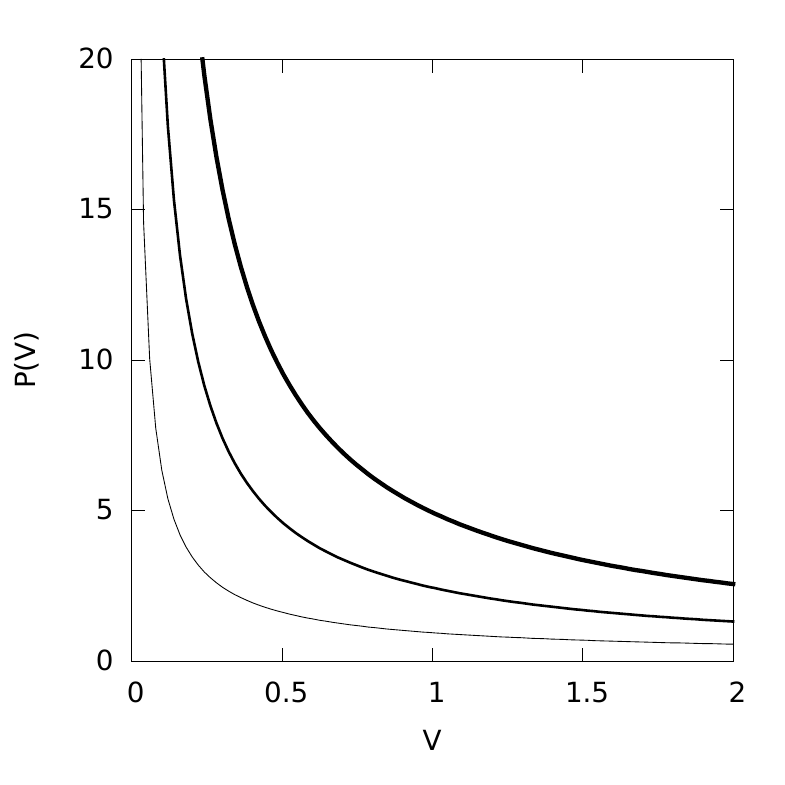}}
\caption{$P(V)$ for linear field with (a) $T=1$ (thin), $T=2$ (medium), $T=3$ (bold), $Q=1$, (b) $Q=1$ (thin), $Q=2$ (medium), $Q=3$ (bold), $T=1$.}
\label{fig_P(V)_PMI_s_1_T_Q}
\end{figure}

The graphs for $P(V)$ are presented on Fig.~\ref{fig_P(V)}. On Fig.~\ref{fig_P(V)_PMI_s_1_T_Q} the dependence of $P(V)$ is demonstrated for linear field for different $T$ and $Q$.

The system of equations for the critical parameters $T_c$ and $V_c$ \cite{kubiz12}
\begin{equation}
\begin{cases}
\left.\left(\frac{\partial P}{\partial V}\right)_{T,Q}\right|_{V=V_c,T=T_c}=0
\\
\left.\left(\frac{\partial^2P}{\partial V^2}\right)_{T,Q}\right|_{V=V_c,T=T_c}=0
\end{cases}
\end{equation}
do not have solutions $T_c>0$, $V_c>0$ for none of the fields. It can also be explained by the monotonous (without of the inflection points) behavior of the function $P=P(V)$ for different $T$ and $Q$ and what is demonstrated on the Fig.~\ref{fig_P(V)} and Fig.~\ref{fig_P(V)_PMI_s_1_T_Q}. Finally we calculate the heat capacity by formula $C=T\left(\frac{\partial S}{\partial T}\right)_{Q,P}$. These results are presented it the Table~\ref{tab_C}, where $W=W\left(\frac{q^2}{\beta^2r_+^2}\right)$. They are expressed in terms of $r_+$, $q$, $\Lambda$ and can be written in terms of thermodynamic variables of $S$, $Q$, $P$. 
\begin{table}[h!]
\begin{center}
\begin{tabular}{|ll|}
\hline
$C=\pi\frac{\Lambda r_++2^{s-1}(2s-1)q^{\frac{2s}{2s-1}}r_+^{-\frac{1}{2s-1}}}{2\Lambda-2^{s}q^\frac{2s}{2s-1}r_+^{-\frac{2s}{2s-1}}}$ & PMI \\
$C=\pi\frac{(\Lambda-2\beta^2)r_++2\beta q\sqrt{1+\frac{\beta^2r_+^2}{q^2}}}{2\Lambda-4\beta^2+\frac{4\beta^3r_+}{q}\left(1+\frac{\beta^2r_+^2}{q^2}\right)^{-\frac{1}{2}}}$ & BI \\
$C=\pi\frac{(\Lambda-2\beta^2)r_+-2\beta^2r_+\operatorname{arsinh}\left(\frac{\beta r_+}{\sqrt{2}q}\right)+2\beta^2r_+\ln\left(\frac{\sqrt{2}\beta r_+}{q}\right)+2\sqrt{2}\beta q\sqrt{1+\frac{\beta^2r_+^2}{2q^2}}}{2\Lambda-4\beta^2\operatorname{arsinh}\left(\frac{\beta r_+}{\sqrt{2}q}\right)+4\beta^2\ln\left(\frac{\sqrt{2}\beta r_+}{q}\right)}$ & LN \\
$C=\pi\frac{(\Lambda+2\beta^2)r_++2\beta q(\sqrt{W}-W^{-\frac{1}{2}})}{2\Lambda+4\beta^2-\frac{4\beta q}{(1+W)r_+}(\sqrt{W}+W^{-\frac{1}{2}})}$ & EN \\
\hline
\end{tabular}
\caption{The heat capacity $C(r_+)$.}
\label{tab_C}
\end{center}
\end{table}

We have for any type of fields there is a value of $r_+$ where $C(r_+)=0$ (as for the temperature) and this point separates two regions, one of them where $C<0$ (for small $r_+$) and the second region where $C>0$ (for large $r_+$). It is known that when $C<0$ a thermodynamic system is unstable, so the black holes are unstable for small $r_+$ (and therefore for small black hole mass $M$), it can be also shown, that the domain where $C<0$ coincides with corresponding domain where $T<0$, as it follows from expression for $C$ and from monotonous behavior of $T(r_+)$ function.

\begin{figure}[h!]
\centering
\includegraphics[width=0.4\textwidth]{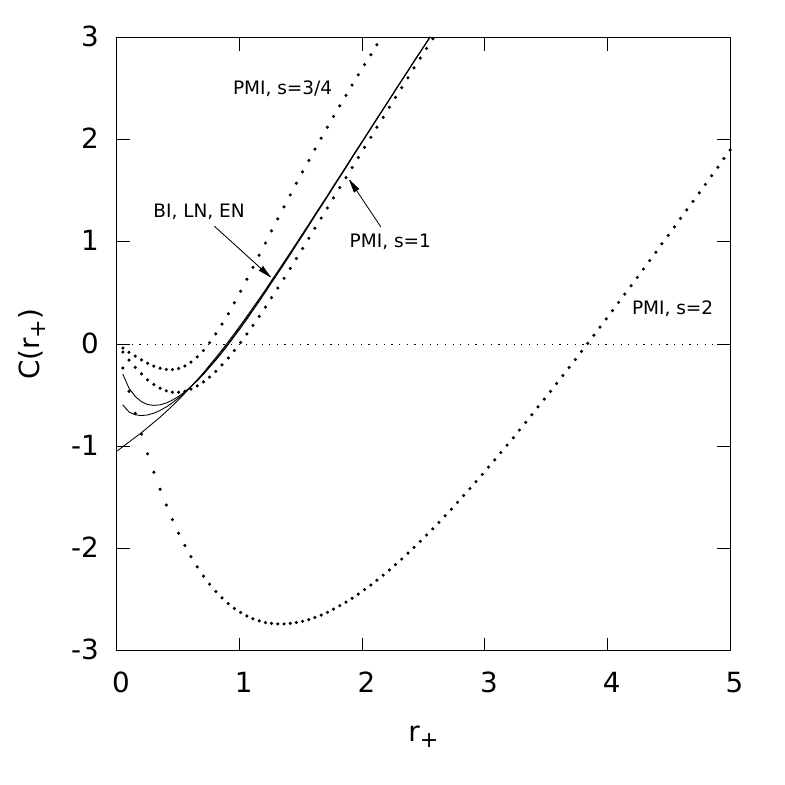}
\caption{The heat capacity $C(r_+)$ for $\beta=1$, $\Lambda=-1$, $q=1$.}
\label{fig_C(r)}
\end{figure}

The graph for $C(r_+)$ are presented on Fig.~\ref{fig_C(r)}. There is a minimum (except for the BI field and this is related to the taken value of $\beta$, for example with $\beta=3$ we get a minimum as for other field types) after which the curves increase monotonously with rise of $r_+$.
\section{Conclusions}
In this paper we considered the system of a static black hole with nonlinear electromagnetic fields and cosmological constant in a three-dimensional (2+1) spacetime. We investigated four kinds of this fields: the PMI, Born-Infeld field, logarithmic field and exponential field. We have obtained exact solutions of field equations and on their basis we have calculated temperature and electric potential on the event horizon of black hole. We derived the first law of black hole thermodynamics in thermodynamic variables of $S$ and $Q$, then we added to it a pressure-volume term $VdP$, and calculated the thermodynamic volume, it is the same for all field types. The first law is also satisfied in this extended case. We obtained equations of state and made a conclusion about nonexistence of critical parameters for all kinds of electromagnetic fields. Also  we calculate the heat capacity. 

The PMI field when $s=1$ describes linear Maxwell electromagnetic field. The PMI field of $s=3/4$ corresponds to the mentioned above conformally invariant case and we considered it here. The Born-Infeld field and logarithmic field allow to eliminate the singularity for radial component of electric field when $r\to0$. The obtained relations for metric function, temperature, pressure and heat capacity for all type of fields show some similarities, namely when $r$ (or $r_+$) becomes large the term with the cosmological constant is dominating, thus for some fixed parameters of the black hole all dependences are very close to each other. The logarithmic and exponential fields also demonstrate some similarities with Born-Infeld case not only for large $r$ ($r_+$), but also for its intermediate values, but for very small values of $r$ ($r_+$) these three types of fields have some discrepancy due to different behavior of the metric function for small $r$. The plotted graphics for mentioned above dependences for all different field types for various values of parameters confirm given above general conclusions. The behavior of curves for PMI field is defined by parameter $s$ and we have consider the case when $s>1/2$, because if $s<1/2$ the electric field is divergent at infinity. For metric function $g(r)$ we have the event horizon whose existence depends on the value of $m\propto M$, where $M$ is the black hole mass. The temperature is the monotonically increasing function of $r_+$. The expressions of $P(V)$ are decreasing functions on $V$ without the existence of phase transition.
\section{Acknowledgements}
This work was partly supported by the projects $\Phi\Phi$-83$\Phi$ (No. 0119U002203), $\Phi\Phi$-63Hp (No. 0117U007190) from the Ministry of Education and Science of Ukraine.

\end{document}